\documentclass[12pt]{article}

\usepackage{graphicx}
\usepackage{graphics}
\usepackage{times}
\usepackage{color}

\newenvironment{natabstract}{%
\begin{quote} \bf}
{\end{quote}}

\title{Noiseless nonreciprocity in a parametric active device}

\author
{Archana Kamal,$^{1}$ John Clarke,$^{2}$ M. H. Devoret$^{1\ast}$\\
\\
$^{1}$Department of Physics and Applied Physics, Yale University,\\
15 Prospect Street, New Haven, CT 06520, USA\\
$^{2}$Department of Physics, University of California,
and Materials Sciences Division,\\
Lawrence Berkeley National Laboratory, Berkeley, CA 94720, USA\\
\\
$^\ast$Correspondence to: M. H. Devoret$^{1}$ email:
michel.devoret@yale.edu. }

\date{}

\begin{document}


\baselineskip24pt


\maketitle

\begin{natabstract}
    Nonreciprocal devices such as circulators and isolators belong to an important class of microwave components employed in applications like the measurement of mesoscopic circuits at
    cryogenic temperatures \cite{RSL1, Saclay, meso1, meso2}. The measurement protocols usually involve an amplification chain which relies on circulators to separate input and output channels
    and to suppress backaction from different stages on the sample under test. In these devices the usual reciprocal symmetry of circuits is broken by the phenomenon of Faraday rotation
    based on magnetic materials and fields \cite{Pozar}. However, magnets are averse to on-chip integration, and magnetic fields are deleterious to delicate
    superconducting devices \cite{magnoise1, magnoise2, magnoise3}. Here we present a new proposal combining two stages of parametric modulation emulating the action of a circulator. It is devoid of magnetic components and suitable for
    on-chip integration. As the design is free of any dissipative elements and based on reversible operation, the device operates noiselessly, giving it an important
    advantage over other nonreciprocal active devices for quantum information processing applications.
\end{natabstract}

Reciprocity is one of the fundamental symmetries frequently
encountered in electrical circuits. It is equivalent to the more
familiar notion of the principle of reversibility in optics which
states that any experiment is symmetric under an exchange of
source and image \cite{opticsbook}. Reciprocity can, however, be
violated, for example, by the magneto-optic effect of Faraday
rotation \cite{Pozar} which leads to rotation of the polarization
vector of light resulting from different propagation velocities of
left- and right-circularly polarized waves in the presence of an
applied magnetic field $\mathbf{B}$ parallel to the direction of
propagation (Fig. 1). The nonreciprocal phenomenon of Faraday
rotation should be contrasted with the superficially similar,
though reciprocal, effect of optical activity where the
polarization vector of light is rotated on passage through a
non-centrosymmetric (chiral) medium. This change in the sense of
rotation of polarization for counterpropagating waves in a Faraday
medium (as seen by the observer receiving the light) has led some
physicists to refer to this effect as a form of
time-reversal-symmetry-breaking, a use of words that we prefer to
avoid here \cite{PTsym, firstnote}.

The phenomenon of nonreciprocity has propelled numerous
theoretical investigations [\cite{NRtheory} and refs. therein];
furthermore it offers immediate practical applications. Recent
progress in solid state superconducting qubits, that provide some
of the most promising architectures for scalable quantum computers
\cite{Nakamura,scqubits}, has generated a huge incentive to
integrate the components required for qubit operations and readout
on-chip for incorporation in future quantum mechanical processors.
A large variety of qubit readout protocols involve microwave
reflection based measurements and rely on nonreciprocal devices
like circulators (or isolators) for separation of input and output
channels \cite{RSL1,Saclay}. These devices also play a strategic
role in measurements based on low-noise microwave parametric
amplifiers which, with the exception of designs based on the
current-biased dc SQUID (Superconducting Quantum Interference
Device), are also operated in a reflection mode with both the
input and output signals collected on the same spatial channel
\cite{castellanos-beltran:083509,yamamoto:042510}. However,
circulators (and isolators) routinely use bulk components made
from ferrites to achieve nonreciprocal phase shifts (Fig. 1c)
through Faraday rotation, making them unamenable to chip
fabrication. Moreover, to bias the magnetic field in the ferrite,
most of these devices use a permanent magnet which may channel
flux into the superconducting device under test.

In this letter, we present the full analysis of a model for a
four-port circulator based on parametric active devices with no
magnetic components. In active devices the energy source --
provided by the pump -- acts as the external ``bias'' field and
sets the reference phase for the system, in analogy with the role
played by the magnetic field in a Faraday medium. We exploit this
effect in a cascade of active devices with pump phases at each
stage tuned appropriately to obtain nonreciprocal transmission.

The main building block of our design is a reversible $IQ$
(in-phase/quadrature) modulator capable of performing noiseless
frequency up- and down-conversion. A convenient analytical model
capturing the fundamental properties of the device is shown in
Fig. 2a. The device comprises two low frequency LC resonators
(addressed by two semi -infinite transmission lines $A$ and $B$)
coupled to a high frequency resonator (addressed by the
transmission line $C$) through time-varying couplings $M_{1},
M_{2}$ that emulate the role of the pump drive in active nonlinear
devices and transfer energy from the tone at $\omega_{c}$ to the
signal modes propagating on the transmission lines. It operates in
a manner analogous to the $IQ$ modulation schemes routinely used
in radiofrequency (RF) communication systems and microwave pulse
engineering (hence the name) and converts two orthogonal spatial
modes travelling on two distinct spatial ports ($A, B$) at same
frequency (here $\omega_{0}$) into two orthogonal temporal modes
travelling on the same spatial line ($C$) at different frequencies
($\omega_{+}, \omega_{-}$). In view of the reversible frequency
conversion performed by this device (Fig. 2a), we will henceforth
refer to it as the up/down-converter (UDC). In practice, such a
device can be implemented on-chip using a ring modulator based on
Josephson junctions, along the lines of the recently demonstrated
experiment with Josephson parametric converter
\cite{bergealJPC,2009arXiv0912.3407B}.

The complete design for the active circulator (Fig. 3a) consists
of a UDC functioning as a frequency up-converter, a phase-shifter
and a second UDC functioning as a frequency down-converter.

A concise representation of the dynamics at each of the three
stages in the cascade is provided by the scattering matrix $S$
which relates the outgoing wave amplitudes to the incoming wave
amplitudes as seen from the ports of a network. We start by
deriving the scattering matrix of the UDC stage. This is done by
evaluating the impedance matrix $Z$ of the UDC, as seen from its
ports, and using the identity \cite{Pozar}
\begin{eqnarray}
    S = (Z + Z_{0})^{-1} \times (Z - Z_{0})
\end{eqnarray}
where
\begin{equation}
Z_{0} = {\rm diag} (Z_{A}, Z_{B}, Z_{C}, Z_{C}), \label{imp}
\end{equation}
with $Z_{A} = Z_{B}$ and $Z_{C}$ denoting the characteristic
impedances of the semi-infinite transmission lines serving as low
and high frequency ports respectively. We obtain (see
supplementary information, Fig. S1)
\begin{equation}
    \left(\matrix{a_{0}^{' \rm out}\cr
    a_{0}^{'' \rm out}\cr
    b_{+}^{\rm out}\cr
    b_{-}^{\dagger \rm out}
    }\right)
= \left(\matrix{
    r_{0} & -q_{0} & t_{d} e^{-i \phi} & s_{d} e^{i \phi} \cr
    q_{0} & r_{0} & i t_{d} e^{-i \phi} & -i s_{d} e^{i \phi} \cr
    t_{u} e^{i \phi} & -i t_{u} e^{i\phi} & r_{+} & 0 \cr
    -s_{u} e^{-i \phi} & - i s_{u} e^{-i\phi} & 0 & r_{-}
    }\right)
\left(\matrix{
    a_{0}^{' \rm in}\cr
    a_{0}^{'' \rm in}\cr
    b_{+}^{\rm in}\cr
    b_{-}^{\dagger \rm in}
    }\right).
    \label{Smatrix}
\end{equation}
Here $a$ and $b$ denote the (reduced) amplitudes or the
annihilation operators for the waves travelling on left and right
transmission lines respectively (see supplementary information for
details). These satisfy bosonic commutation relations of the form
\cite{IOT}
\begin{equation}
    [a_{i}, a^{\dagger}_{j}] = \delta (\omega_{i} -\omega_{j}).
    \label{IOTcomm}
\end{equation}
In writing Eqs. (\ref{Smatrix}) and (\ref{IOTcomm}), we have set
$a_{0} = a [\omega_{0}], a_{+} = a[\omega_{+}], a_{-} =
a[\omega_{-}]$ (see Fig. \ref{Fig2}b). Similarly, the reflection
coefficients at various ports are denoted by $r_{0}, r_{+}$ and
$r_{-}$. The cross reflection between the low frequency signal
ports is denoted by $q_{0}$. The transmission coefficients are
written as $t$ (transmission without conjugation) and $s$
(transmission with conjugation) with subscripts $(u, d)$
indicating the up-conversion and down-conversion respectively. It
is useful to note that the phase of the carrier, denoted by
$\phi$, affects only the transmitted amplitudes and rotates the
two sidebands in opposite directions as can be seen from the
corresponding scattering coefficients $s$ and $t$ in Eq.
(\ref{Smatrix}). The invariance of reflection amplitudes to the
phase of the coupling will be important in understanding total
reflections of the cascade, as we describe later.

Further, we note that the matrix obtained in Eq. (\ref{Smatrix})
is non-unitary, that is $S^{\dagger} S \neq 1$, which implies
nonconservation of photon number as is natural for an active
device. The matrix recovers its unitary form as we turn off the
couplings $M_{1}, \;M_{2}$ responsible for energy transfer between
the pump and the signal modes. The full $8 \times 8$ matrix
[supplementary information, Eq. (S18)], describing the device
operation for all modes and their respective conjugates, fulfils
the fundamental requirement of
\emph{symplecticity}\cite{bergealJPC}.

We can similarly describe the action of the frequency-independent
phase shifting (PS) stage using a scattering matrix of the form
\begin{equation}
    \left(\matrix{a_{+}^{\rm out}\cr
    a_{-}^{\dagger \rm out}\cr
    b_{+}^{\rm out}\cr
    b_{-}^{\dagger \rm out}
    }\right)
     = \left(\matrix{
    0 & 0 & e^{-i\theta} & 0 \cr
    0 & 0 & 0 & e^{i\theta} \cr
    e^{-i\theta} & 0 & 0 & 0 \cr
    0 & e^{i\theta} & 0 & 0
    }\right)
    \left(\matrix{a_{+}^{\rm in}\cr
    a_{-}^{\dagger \rm in}\cr
    b_{+}^{\rm in}\cr
    b_{-}^{\dagger \rm in}
    }\right).
    \label{PS}
\end{equation}

For each stage of the cascade, we now go from the scattering
matrix representation to the transfer matrix representation
\cite{Pozar},
\begin{equation}
    \left(\matrix{a_{+}^{\rm out}\cr
    a_{-}^{\dagger \rm out}\cr
    b_{+}^{\rm out}\cr
    b_{-}^{\dagger \rm out}
    }\right)
    = S
    \left(\matrix{a_{+}^{\rm in}\cr
    a_{-}^{\dagger \rm in}\cr
    b_{+}^{\rm in}\cr
    b_{-}^{\dagger \rm in}
    }\right)
    \quad
    \mapsto
    \quad
    \left(\matrix{b_{+}^{\rm out}\cr
    b_{+}^{ \rm in}\cr
    b_{-}^{\dagger \rm out}\cr
    b_{-}^{\dagger \rm in}
    }\right)
    = T
    \left(\matrix{a_{+}^{\rm in}\cr
    a_{+}^{\rm out}\cr
    a_{-}^{\dagger \rm in}\cr
    a_{-}^{\dagger \rm out}
    }\right),
    \label{transfer}
\end{equation}
since it is straightforward to calculate the total transfer matrix
of the device by multiplying the respective transfer matrices of
different stages \cite{thirdnote},
\begin{equation}
    T_{\rm total}= T_{DC_{R}} \times T_{PS} \times T_{UC_{L}}.
\end{equation}
Here the subscripts $L, R$ index the left hand upconversion ($UC$)
and right hand downconversion ($DC$) stage (Fig. \ref{Fig3}a). The
scattering matrix of the whole device is then obtained from
$T_{\rm total}$ using the inverse of the transformation in Eq.
(\ref{transfer}) (see supplementary information). We also note
that
\begin{eqnarray}
    T_{DC} &=& F^{-1} \times T_{UC}^{-1} \times F \nonumber\\
                 &=& F \times T_{UC}^{-1} \times F, \;\;(F^{-1}=F)
\end{eqnarray}
where $F = \sigma_{X} \otimes I_{2}$, ($\sigma_{X}$ is the 2D
pauli spin matrix and $I_{2}$ is the 2D unity matrix). This matrix
$F$ is required to flip the indices, thus maintaining consistency
in labelling the `in' and `out' amplitudes along a given direction
of propagation.

In our analysis we consider the operation at resonance, that is,
when the input signal frequency coincides with the band center of
the input resonators. Setting the phase of the pump at the first
UDC stage $\phi_{L} = 0$ for calculational simplicity, we observe
a transmission resonance for $\theta = \pm \pi/2$ (phase rotation
by the PS stage), $\phi_{R} =\pi/4$ (phase of the pump at the
second UDC stage), $\delta_{\pm} =1/\sqrt{2}$ (detuning of the
sidebands from the carrier in units of linewidth i.e. half width
at full maximum of the resonance lineshape) of the high frequency
resonator), and $\alpha_{L} = \alpha_{R} = M_{0}/\sqrt{L_{A,B}
L_{C}} = 2^{-3/4}$ (strength of the parametric coupling). For this
choice of parameters, we obtain the scattering matrix of the
complete device as
\begin{equation}
 S_{\rm total} = \left(\matrix{
    0 & 0 & 0 & i\cr
    0 & 0 & -i & 0\cr
    i & 0 & 0 & 0\cr
    0 & i & 0 & 0
 }\right).
 \label{final}
\end{equation}
This is the matrix of a perfect four-port circulator. The analogy
between a conventional circulator and the active circulator design
proposed in this paper is made apparent from the respective wave
propagation diagrams in Figs. 1c and 3b (see supplementary
information for details on the calculation of coefficients on
different arms in 3b). Nonetheless there are important differences
between the two designs despite the identity of the final $S$
matrix. The coefficients on the forward (green) and backward (red)
propagating arms of the active circulator design (Fig. 3b) involve
deamplification followed by amplification, unlike the passive
splitters (90 or 180 degree hybrids) employed in Faraday rotation
schemes. This can be observed by squaring the amplitudes on each
of the two arms originating from (or terminating into) a port and
calculating the net power output, for each isolated UDC stage. It
is straightforward to observe that, unlike the case of Fig. 1c,
they do not add up to unity. Nonetheless, the overall transmission
is unity due to an exact cancellation of the reduction and gain in
amplitudes. The wave propagation diagrams in Fig. 3b reveal
another important difference of this design from that of a
conventional circulator. The non-reciprocal action of the active
circulator is not based upon any non-reciprocal phase shifters;
instead it relies on the active stages used for frequency up- and
down-conversion. The phase matching condition in the forward
direction is met by tuning the phase of the coupling at the input
and output UDC stages. In the reverse direction the phase mismatch
leads to unity transmission in the spatially orthogonal port
instead, leading to complete isolation between the incident signal
port and its corresponding output port.

Fig. \ref{Fig3}c shows a convenient method to visualize this
circulator action geometrically by mapping the device dynamics at
different stages using a modulation ellipse. This approach is
inspired by the polarization ellipse used to represent of state of
polarization of an electromagnetic wave (linear, circular or
elliptical), which involves recording the trajectory traced out by
the tip of the polarization vector of light (defined by the
instantaneous direction of the electric field vector $\textbf{E}$)
in a plane perpendicular to the direction of propagation.
Equivalently, a two-dimensional representation of the components
$E_{X}$ and $E_{Y}$ of the electric field in the complex plane can
be used to obtain a geometric description of the polarization of
the light wave. In the case of a modulation ellipse representation
of the dynamics of the proposed device, we extend this idea to map
\emph{two} distinct orthogonal modes (x,y) at each stage of the
device [spatial: ($x=a_{0}^{'}, y=a_{0}^{''}$) or temporal:
($x=a_{+}, y=a_{-}$)] as an ellipse in the plane defined by the
coordinates $I= Re[x+y], Q =Im[x-y^{*}]$. This exercise shows that
the final ellipses obtained at the output in case of forward and
backward propagation through the device are rotated by $90$
degrees with respect to each other. This indicates that in the
case of reverse propagation the orthogonal spatial port, relative
to the forward propagation, receives the transmitted energy
leading to a circulator action (see supplementary information and
Fig. S2 for more details).

Furthermore, as seen from Eq. (\ref{Smatrix}), the reflection
coefficients at the UDC stages are non-zero for all modes.
However, for the whole cascade, the total reflection is
identically zero at every port [$s_{ii} = 0 \; {\rm for \; all}\;
i$ in Eq. (\ref{final})]. This remarkable cancellation of total
reflections for the cascade can be understood in analogy with a
Fabry-Perot resonance where a cavity flanked by two identical
reflecting mirrors displays unity transmission at resonance. The
total phase shift between the active ``mirrors'' in our device:
$(\pi/2)_{a_{+}} - (- \pi/2)_{a_{-}^{\dagger}} = \pi$, is akin to
the resonance condition when a half-wavelength of the incident
radiation equals the length of the Fabry-Perot cavity. Also the
reflections at the two UDC stages are identical [as the reflection
coefficients are independent of the phase angle $\phi$, cf. Eq.
(\ref{Smatrix})], fulfilling the second condition for the
transmission resonance and net cancellation of reflections
\cite{fourthnote}.

We show the dependence of circulator action on different
parameters in the device in Fig. 4. Since the isolation achieved
is robust to reasonable deviations of parameters from their ideal
values ($\phi_{L}=0, \;\phi_{R} = \pi/4, \;\theta =\pi/2,
\;\delta_{\pm} =1/\sqrt{2}, \;\alpha_{L} = \alpha_{R} =2^{-3/4}$),
the active circulator design holds promise for use in practical
circuits. Another interesting feature of this device is the
reversal of transmission characteristics with phase of the pumps
($\phi_{L,R} \mapsto - \phi_{L, R}$) (Fig. 4c). In the classic
circulators based on passive Faraday rotation, this can be
accomplished by changing the polarity of the magnetic bias field.
Thus the clock (``pump phase'') in an active device indeed plays a
role equivalent to the magnetic field in a Faraday medium.

In conclusion, we have described a scheme for achieving
nonreciprocal wave propagation using a protocol involving
up-conversion followed by down-conversion mediated by an
appropriate phase shift \cite{secondnote}. The proposed design
performs noiselessly as it consists of purely dispersive
components with no dissipation, making it attractive for quantum
information applications using superconducting circuits
\cite{RSLQMP}. Besides microwave applications, the architecture of
the protocol described in this paper can be adapted to optical
frequencies, where it can complement the recently proposed designs
of nonreciprocal light propagation based on dynamical modulation
of the refractive index of photonic structures \cite{natphoton}
and the use of a surface waveguide on photonic crystals
\cite{liu:021119}. In addition to the practical applications
outlined above, the treatment described in this letter may also
give theoretical insights into the inherently directional dynamics
of devices like the dc SQUID \cite{SQUIDhandbook1}, when
additional active stages are included in the chain. This can be
useful in tackling unanswered questions pertaining to the quantum
noise of dc SQUID amplifiers \cite{SQUIDhandbook2}.

\bibliographystyle{unsrt}

\begin{thebibliography}{10}

\bibitem{RSL1}
{Wallraff, A. \it{et al.}}
\newblock Strong coupling of a single photon to a superconducting qubit using
  circuit quantum electrodynamics.
\newblock {\em Nature}, 431:162, 2004.

\bibitem{Saclay}
{Mallet, F. \it{et al.}}
\newblock Single-shot qubit readout in circuit quantum electrodynamics.
\newblock {\em Nat Phys}, 5:791--795, 2009.

\bibitem{meso1}
{Johnson, B. R. \it{et. al.}}
\newblock Quantum non-demolition detection of single microwave photons in a
  circuit.
\newblock {\em Nat. Phys.}, 6:1745--2473, 2010.

\bibitem{meso2}
{Naaman, O., Aumentado, J., Friedland, L., Wurtele, J. S. \&
Siddiqi, I.}
\newblock Phase-locking transition in a chirped superconducting josephson
  resonator.
\newblock {\em Phys. Rev. Lett.}, 101(11):117005, 2008.

\bibitem{Pozar}
{Pozar, David M.}
\newblock {\em Microwave Engineering}, pages 471--482.
\newblock Wiley, ed. 3, 2005.

\bibitem{magnoise1}
{Van Harlingen, D. J. \it{et. al.}}
\newblock {Decoherence in Josephson-junction qubits due to critical-current
  fluctuations}.
\newblock {\em Phys. Rev. B}, 70(6):064517, 2004.

\bibitem{magnoise2}
{Kakuyanagi, K. \it{et. al.}}
\newblock Dephasing of a superconducting flux qubit.
\newblock {\em Phys. Rev. Lett.}, 98(4):047004, 2007.

\bibitem{magnoise3}
{Choi, S., Lee, D.-H., Louie, S. G. \& Clarke, J.}
\newblock Localization of metal-induced gap states at the metal-insulator
  interface: Origin of flux noise in squids and superconducting qubits.
\newblock {\em Phys. Rev. Lett.}, 103(19):197001, 2009.

\bibitem{opticsbook}
{Schuster, A.}
\newblock {\em An Introduction to the Theory of Optics}, pages 41--45.
\newblock Edward Arnold, London, 1904.

\bibitem{PTsym}
{Barron, L. D.}
\newblock {Parity and Optical Activity}.
\newblock {\em Nature}, 238:17--19, 1972.

\bibitem{firstnote}
This confusion can be resolved by considering the symmetries of
$\mathbf{E}$
  and $\mathbf{B}$ fields under parity ($P$) and time reversal ($T$)
  operations. The electric field $E$ described by a \emph{polar vector} that is
  \emph{symmetric under time reversal} and the magnetic field $B$ described by
  an \emph{axial vector} (more appropriately as an antisymmetric tensor of rank
  two) that is \emph{anti-symmetric under time reversal} lead to the invariance
  of Maxwell's equations describing the electromagnetic field under both $P$
  and $T$ operations. Thus any physical process involving only electromagnetic
  interactions also preserves these two symmetries. In other words, if $P$ and
  $T$ are applied to a \emph{complete} experiment, the resulting situation
  should also be physically realizable as a solution of Maxwell's equations. It
  can be checked that Faraday rotation under both $P$ and $T$ operations leads
  to feasible outcomes, thus preserving both the symmetries.

\bibitem{NRtheory}
{Potton, R. J.}
\newblock Reciprocity in optics.
\newblock {\em Rep. Prog. Phys.}, 67:717--754, 2004.

\bibitem{Nakamura}
{Ladd, T. D. \it{et al.}}
\newblock Quantum computers.
\newblock {\em Nature}, 464:45--53, 2010.

\bibitem{scqubits}
{Clarke, J. \& Wilhelm, F. K.}
\newblock Superconducting quantum bits.
\newblock {\em Nature}, 453:1031--1042, 2008.

\bibitem{castellanos-beltran:083509}
{Castellanos-Beltran, M. A. \& Lehnert, K. W.}
\newblock Widely tunable parametric amplifier based on a superconducting
  quantum interference device array resonator.
\newblock {\em Appl. Phys. Lett.}, 91(8):083509, 2007.

\bibitem{yamamoto:042510}
{Yamamoto, T. \it{et al.}}
\newblock Flux-driven josephson parametric amplifier.
\newblock {\em Appl. Phys. Lett.}, 93(4):042510, 2008.

\bibitem{bergealJPC}
{Bergeal, N \it{et al.}}
\newblock Analog information processing at the quantum limit with a josephson
  ring modulator.
\newblock {\em Nat Phys}, 6:296--302, 2010.

\bibitem{2009arXiv0912.3407B}
{Bergeal, N. \it{et al.}}
\newblock {Phase preserving amplification near the quantum limit with a
  Josephson Ring Modulator}.
\newblock available at http://arxiv.org/abs/arXiv:0912.3407.

\bibitem{IOT}
{Yurke, B.}
\newblock Input output theory.
\newblock In {Drummond, P.D. \& Ficek, Z.}, editor, {\em Quantum Squeezing},
  pages 53--95. Springer, 2004.

\bibitem{thirdnote}
We note that for the case of $IQ$ coupling considered in Fig.
2(A), the terms
  coupling the two sidebands, $s_{34}$ and $s_{43}$, are zero [Eq.
  (\ref{Smatrix})]. If the phase difference between the two couplings $M_{1}$
  and $M_{2}$ deviates from 90 degrees, crosstalk appears between the two
  sidebands generated at port C. Furthermore, if $M_{1}$ and $M_{2}$ would be
  completely in phase, the two sidebands would be maximally coupled while the
  cross reflections $q_{0}$ between the low frequency input ports would reduce
  to zero. In such a case, the transformation defining the transfer matrix is
  singular (refer to methods under supplementary information for details).

\bibitem{fourthnote}
A point of disticntion between the two pictures is that the
symmetry of the
  active circulator design is described by a sub-group of the $SU(4)$ group
  (the group formed by $4\times4$ complex matrices of unit determinant, Eq.
  \ref{final}]), and not by the $SU(2)$ group that describes passive lossless
  two-port devices like the Fabry-Perot resonator.

\bibitem{secondnote}
A recent theoretical work \cite{Jens} showed the existence of a
non-reciprocal effect using nonlinear circuit that involves
microwave resonators coupled through Josephson junctions. While
the proposal of this reference is based on a passive Josephson
circuit, it involves, unlike our proposal, very small JJ
susceptible to offset charges. Furthermore, due to its  small
characteristic energy, that device will handle the qubit readout
signals with a lesser throughput than our proposed device.

\bibitem{RSLQMP}
{DiCarlo, L. \it{et. al.}}
\newblock {Demonstration of two-qubit algorithms with a superconducting quantum
  processor}.
\newblock {\em Nature}, 460:240--244, 2009.

\bibitem{natphoton}
{Yu, Z. \& Fan, S.}
\newblock Complete optical isolation created by indirect interband photonic
  transitions.
\newblock {\em Nature Photonics}, 3:91--94, 2009.

\bibitem{liu:021119}
{Liu, A. Q., Khoo, E. H., Cheng, T. H., Li, E. P. \& Li, J.}
\newblock A frequency-selective circulator via mode coupling between surface
  waveguide and resonators.
\newblock {\em Appl. Phys. Lett.}, 92(2):021119, 2008.

\bibitem{SQUIDhandbook1}
{Clarke, J. \& Braginsky, A. I.}, editor.
\newblock {\em The SQUID Handbook Vol. I Fundamentals and Technology of SQUIDs
  and SQUID Systems}.
\newblock Wiley-VCH Verlag GmbH and Co. KGaA, 2004.

\bibitem{SQUIDhandbook2}
{Clarke, J. \& Braginsky, A. I.}, editor.
\newblock {\em The SQUID Handbook Vol. II Applications of SQUIDs and SQUID
  Systems}.
\newblock Wiley-VCH Verlag GmbH and Co. KGaA, 2006.

\bibitem{Jens}
{Koch, J., Houck, A. A., Le Hur, K. \& Girvin, S. M.}
\newblock {Time-reversal symmetry breaking in circuit-QED based photon
  lattices}.
\newblock {\em arXiv:1006.0762}, 2010.

\end{thebibliography}

\subsection*{Acknowledgements} We acknowledge useful discussions
with S. M. Girvin, Jens Koch and R. J. Schoelkopf. This research
was supported by the US National Security Agency through the US
Army Research Office grant W911NF-05-01-0365, the W. M. Keck
Foundation, the US National Science Foundation through grant
DMR-032-5580 (A.K. and M.H.D.) as well as by the Director, Office
of Science, Office of Basic Energy Sciences, Materials Sciences
and Engineering Division, of the U.S. Department of Energy under
Contract No. DE-AC02- 05CH11231 (J.C.). M.H.D. also acknowledges
partial support from the College de France and from the French
Agence Nationale de la Recherche.

\subsection*{Author Contributions} M.H.D. proposed the original
idea for the device. A.K. developed the ideas and performed the
theoretical analysis. J.C. contributed extensively to the
discussions of the results. All authors contributed in writing the
manuscript.

\begin{figure}
  \centering
  \includegraphics[width=0.65\textwidth, height=0.6\textheight]{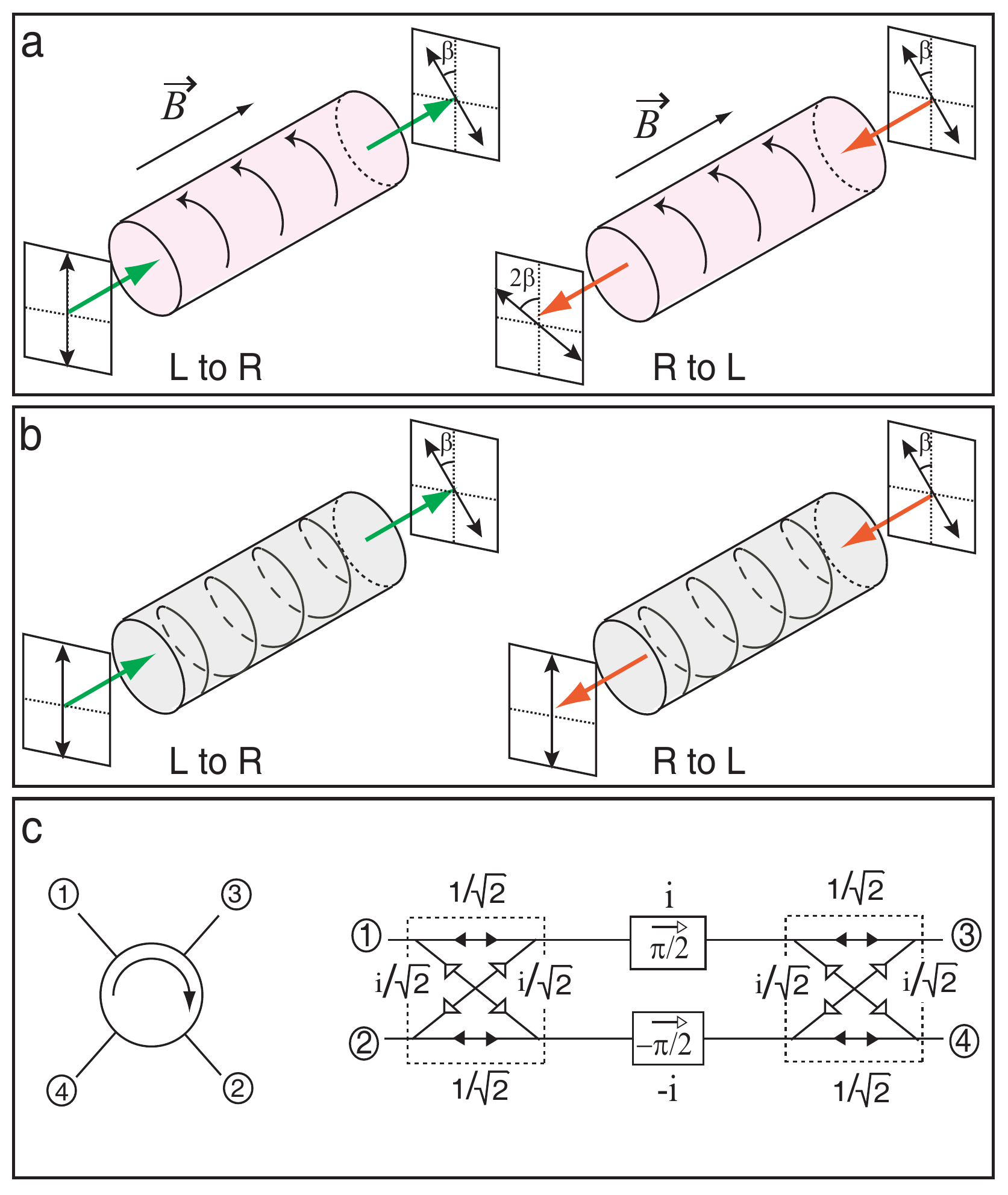}\\
  \caption{{\bf Faraday rotation and circulator action.} {\bf a,} Faraday rotation for a wave travelling from
  left to right in a Faraday-active medium, followed by a reflection back into the medium leading to a reversal of the direction of propagation. The
  rotation of the light polarization is fixed to a rotation-like property of the medium (shown by the arrows), set by an external magnetic
  field oriented along the propagation axis. The sense of light rotation as seen with respect to
  the direction of propagation remains the same, leading to the doubling of the
  rotation angle on reversing the ray through the medium. {\bf b,} Rotation of the polarization vector of light on passage through an
  optically active medium, on the other hand, cancels out on reversing the
  direction of propagation. This occurs because optical rotation depends on the chirality of the medium (represented as a helix)
  which also reverses with the direction of propagation.
  {\bf c,} Representation and schematic design of a conventional four-port circulator. The device consists of two 90 degree hybrids (equivalent to optical beam
  splitters) separated by a \emph{non-reciprocal} phase shifter based on Faraday
  rotation. Solid black arrows indicate an amplitude split with no phase change while, open arrows indicate an amplitude
  split with a 90 degree phase change.
  The non-reciprocal phase shift is effective only for the
  propagation direction indicated by the arrow on the phase shifter
  box.
  }\label{Fig1}
\end{figure}

\begin{figure}
  \centering
  \includegraphics[width=0.8\textwidth]{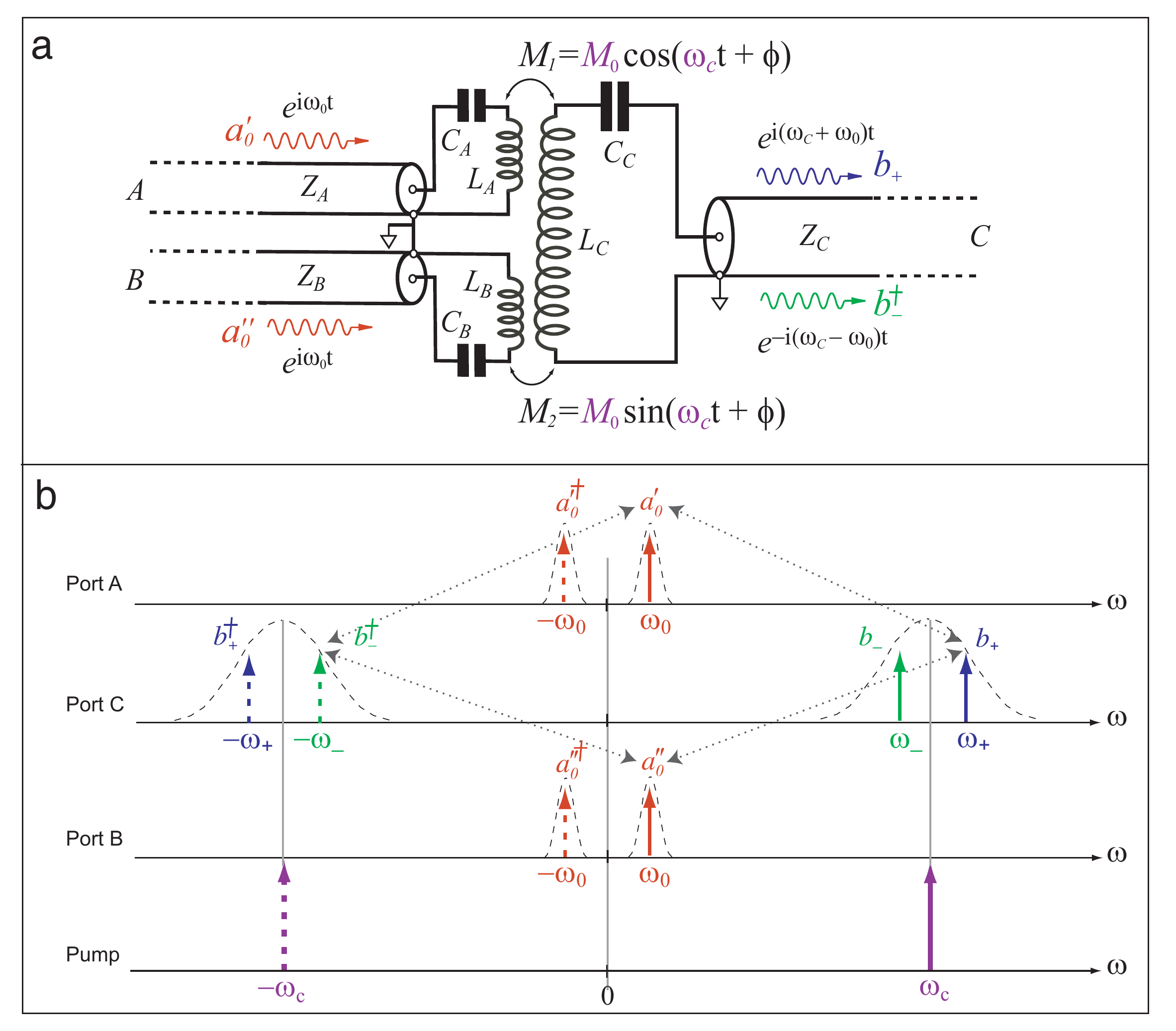}\\
  \caption{{\bf Description of an active reversible (information-conserving) $IQ$ modulator
  performing frequency up- and down-conversion (UDC).} {\bf a,} Circuit schematic of the UDC containing only dispersive
  components.  The two low frequency series LC resonators (with $L_{A} = L_{B}$ and $C_{A} = C_{B}$), are fed by two input semi-infinite transmission lines,
  $A$ and $B$, and parametrically coupled to a third high frequency series LC resonator leading to an output line $C$. The parametric coupling is
  achieved by varying the mutual inductances $M_{1}$ and $M_{2}$ between the left and right resonators at the carrier frequency
  $\omega_{c}$ which, for optimal frequency conversion, is set at the band center of the right
  resonator. When operated from
  left to right, the circuit performs the modulation of low frequency signals
  of frequency $\omega_{0}$ travelling on ports $A$ and $B$ to generate sidebands at $\omega_{c} \pm \omega_{0}$
  travelling on the high frequency line $C$. It performs the inverse operation of demodulation when operated in reverse from right to left.
  {\bf b,} Spectral density/response landscape for different spatial channels (or ports) of the
  UDC circuit in {\bf a} as a function of frequency. The dotted lines represent the couplings between different ports.
  The solid and the dashed arrows represent different frequencies and respective conjugates. The resonance lineshapes of the two spatially distinct
  ports $A$ and $B$ are centered at $\omega_{A} = \omega_{B} = 1/\sqrt{L_{A,B}C_{A,B}}$. Here we show the case when the incoming signal at $\omega_{0}$
  is resonant with the center frequency ($\omega_{0}=\omega_{A,B}$). The two sidebands generated by the UDC on channel $C$ are detuned from the carrier
  $\omega_{c}$ by equal amounts.
  }\label{Fig2}
\end{figure}

\begin{figure}
  \centering
  \includegraphics[width=\textwidth]{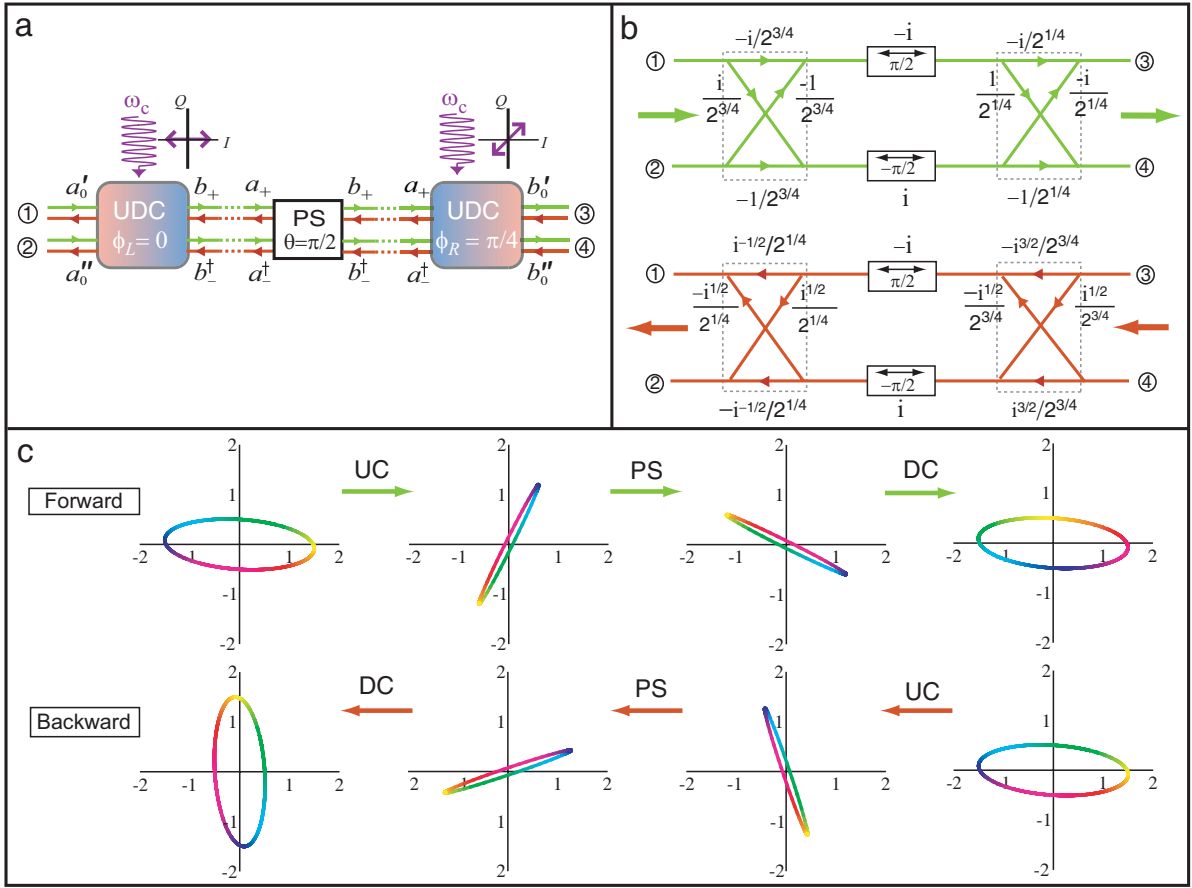}\\
  \caption{
  {\bf Description of the active circulator.} {\bf a,} Circuit schematic of the active circulator design: the first UDC stage acts as a frequency up-converter (UC)
  (also indicated by a gradation in the color of the relevant box) with a parametric coupling modulated at the carrier frequency $\omega_{c} =\omega_{+} - \omega_{0} =
  \omega_{-} + \omega_{0}$ and a phase $\phi_{L} = 0$. This is followed by a phase shifter (PS) that phase shifts both
  the sidebands by $\pi/2$, in opposite directions. They are then demodulated by the final UDC stage acting as a frequency down-converter (DC), with the carrier phase $\phi_{R} = \pi/4$.
  {\bf b,} Forward (green) and backward (red) propagation diagrams calculated using transfer
  matrix method for {\bf a} with appropriate choice of detuning ($\delta_{\pm} =1/\sqrt{2}$) and coupling strengths ($\alpha_{L} = \alpha_{R} = 2^{-3/4}$) for maximum
  isolation.
  {\bf c,} Representation of the device operation using modulation
  ellipses at each stage in the cascade. The top panel shows the forward propagation
  for the case when two distinct signals enter ports 1 and 2 respectively
  ($a_{0}^{'}= 1, a_{0}^{''} =0.5 e^{i \eta}$; $\eta = \pi/24$) while the bottom panel shows the backward
  propagation dynamics when the same signals enter ports 3 and 4
  respectively ($b_{0}^{'}= 1, b_{0}^{''} =0.5 e^{i \eta}$). The relative phase and amplitudes are chosen to represent the most general case
  of two input signals which differ in both amplitude and phase.  The relative phase difference between the two signals is encoded
  as the tilt of the modulation ellipse in the $IQ$ plane while their average phase is represented as
  the color along the perimeter of the ellipse with yellow
  indicating zero phase (also see supplementary information for more details).
}\label{Fig3}
\end{figure}

\begin{figure}
  \includegraphics[width=\textwidth]{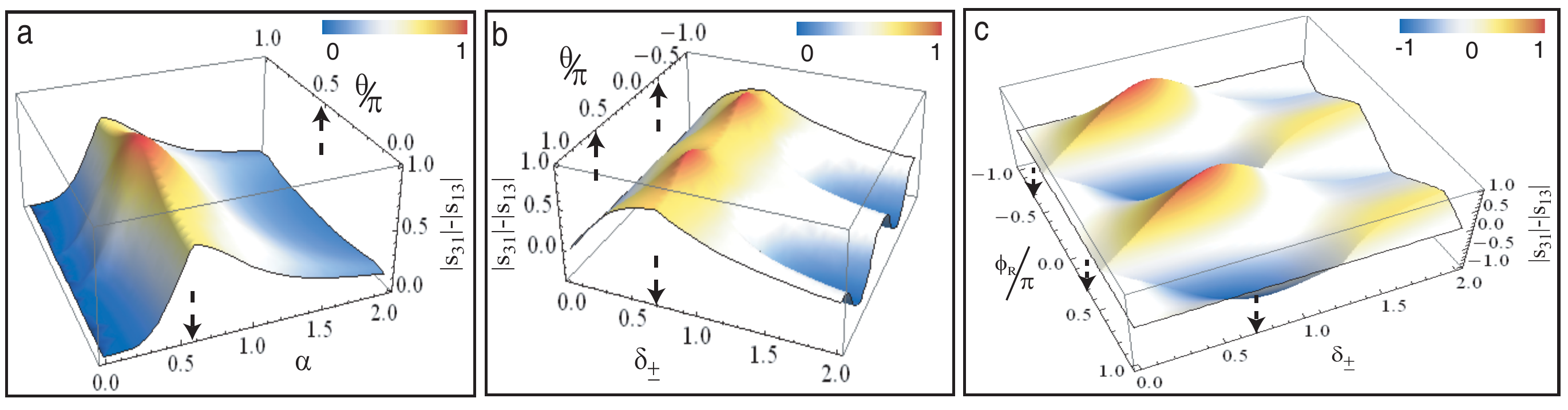}\\
  \caption{{\bf Variation of the difference between forward and backward
  transmission coefficients ($|s_{31}|- |s_{13}|$).}  Asymmetry in transmission, calculated for coupling angles
  $\phi_{L}=0$ and $\phi_{R} = \pi/4$, as a function of {\bf a,} strength of
  the coupling $\alpha$ and phase rotation $\theta$ performed by the
  second phase shifting stage, {\bf b,} detuning $\delta_{\pm}$ of the sidebands from the carrier and phase rotation $\theta$, and {\bf c,} detuning $\delta_{\pm}$
  and the phase of the pump at second UDC stage $\phi_{R}$.
  The points of maxima correspond to the ideal values reported in the text. The plot in {\bf b}, also shows the
  periodicity of the response of the device as a function of $\theta$. In {\bf c}, the variation with respect to the pump phase shows the reversal of transmission
  characteristics with $\phi_{R} \rightarrow -\phi_{R}$. As in {\bf b}, the response is periodic in $\phi_{R}$ with a
  period equal to $\pi$. It can be seen that the design continues to work for moderate
  deviations from the preferred phase angle $\theta = \pi/2$, coupling $\alpha_{L,R} = 2^{-3/4}$, detuning $\delta_{\pm} = 1/\sqrt{2}$
  and pump phase $\phi_{R} = \pi/4$ (values indicated with dashed arrows along the axis).
  }\label{Fig4}
\end{figure}
\end{document}



\baselineskip24pt


\maketitle

\section*{Methods}

\paragraph*{Derivation of scattering matrix.}
In this section, we present the details of the derivation of the
scattering matrix $S$ for the $UDC$ stage. A scattering matrix has
the general form
\begin{equation}
    \vec{a}[\omega]^{\rm out} = S [\omega, \omega^{'}]\vec{a}^{\rm
    in}[\omega^{'}],
    \label{Srelation}
\end{equation}
where  $\vec{a}^{\rm in/out}$ is a column vector formed by the
reduced mode amplitudes (in terms of photon number at the relevant
frequencies) of the network. This description extends to all the
spatial \emph{and} temporal modes of the network. A convenient and
simple way of deriving the $S$ matrix involves calculating the
impedance matrix $Z$ using usual circuit theory, and then using
the identity \cite{Pozar}
\begin{equation}
    S = (Z + Z_{0})^{-1}.(Z- Z_{0})
    \label{ZtoS}
\end{equation}
to obtain the $S$ matrix. Here $Z_{0}$ denotes the characteristic
impedance of the transmission lines acting as channels for
propagation of the incoming and outgoing signals at various ports.
It is of the form
\begin{equation}
Z_{0} = {\rm diag} (Z_{A}, Z_{B}, Z_{C}, Z_{C}), \label{imp}
\end{equation}
where $Z_{A} = Z_{B}$ and $Z_{C}$ are characteristic impedances of
the semi-infinite transmission lines addressing the low and high
frequency ports respectively.

For the parametrically coupled series LC oscillators forming the
UDC stage, Fig. S1a, we obtain the total impedance matrix $Z$ by
adding the inductive ($Z_{\rm ind}$) and capacitive ($Z_{\rm
cap}$) contributions respectively.
\begin{figure}
\centering
  \includegraphics[width=0.8\textwidth]{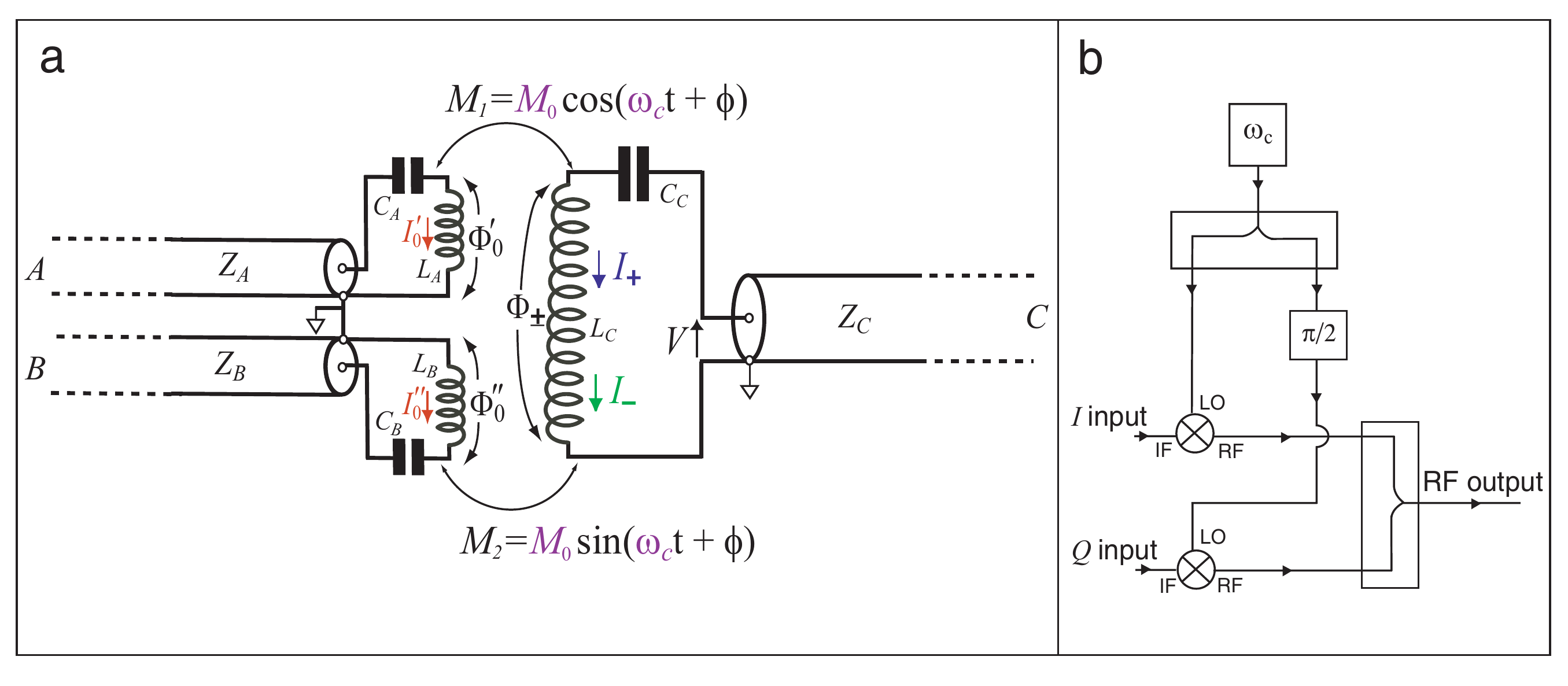}\\
  \caption{{\bf Comparison of two $IQ$ modulation schemes:} {\bf a} Detailed circuit schematic of the $UDC$ performing reversible $IQ$ modulation.
  The fluxes across different inductances denoted as $\Phi_{i}$ and the respective mode
  currents denoted as $I_{i}$ are shown for each series LC resonator. The mutual inductances $M_{1}, \; M_{2}$ lead to off-diagonal coupling (mixing) between the current
  of the left resonators and the flux across the inductance of the right resonator and vice versa.
  {\bf b,} Block diagram of a non-reversible circuit performing the operation described in {\bf a}. The circuit employs dissipative components like
  the mixers (represented by the cross in circle symbol). The CW tone from the generator imposing a carrier frequency $\omega_{c}$ is split into two
  copies, one phase shifted by $\pi/2$ with respect to the other. The carrier is modulated by the signal at $\omega_{0}$ and encoded on two separate
  channels: the `in phase" (or $I$) component and the ``quadrature" (or $Q$) component. The $I$ and $Q$ channels are then combined to
  propagate onto a single spatial channel.}\label{FigS1}
\end{figure}
The inductance matrix $L$ defines the constitutive relationship
between the currents and fluxes for different inductances of the
circuit:
\begin{eqnarray}
    \left(\matrix{
        \Phi_{0}^{'} (t)\cr
        \Phi_{0}^{''} (t)\cr
        \Phi_{+} (t) \cr
        \Phi_{-}(t) \cr
    }\right)
    &=& \left(\matrix{
        L_{A} & 0 & M_{0} e^{-i\phi} & M_{0} e^{i\phi}\cr
        0 & L_{B} & i M_{0} e^{-i\phi} & -i M_{0} e^{i\phi}\cr
        M_{0} e^{i\phi} & -i M_{0} e^{i\phi} & L_{C} & 0\cr
        M_{0} e^{-i\phi} & i M_{0} e^{-i\phi} & 0 & L_{C}
    }\right)
    \left(\matrix{
        I_{0}^{'} (t)\cr
        I_{0}^{''} (t)\cr
        I_{+} (t)\cr
        I_{-} (t)\cr
    }\right) \\
    &=& L
    \left(\matrix{
        I_{0}^{'}(t)\cr
        I_{0}^{''}(t)\cr
        I_{+} (t)\cr
        I_{-} (t)\cr
    }\right).
\end{eqnarray}
In writing the above matrix, we ignore the fluxes at higher
harmonics of the modes at $\omega_{0}$ and $\omega_{\pm}$. The
inductive contribution to the impedance is then calculated by
using the identity
\begin{equation}
\matrix{
    & \left(\matrix{
        V_{0}^{'}\cr
        V_{0}^{''}\cr
        V_{+}\cr
        V_{-}\cr
    }\right)
    =
    \left(\matrix{
        \dot{\Phi}_{0}^{'}\cr
        \dot{\Phi}_{0}^{''}\cr
        \dot{\Phi}_{+}\cr
        \dot{\Phi}_{-}\cr
    }\right)
    = L
    \frac{d}{dt}
        \left(\matrix{
        I_{0}^{'}\cr
        I_{0}^{''}\cr
        I_{+}\cr
        I_{-}\cr
    }\right) \cr
    \cr
\Rightarrow & V[\omega] = Z_{\rm ind} I [\omega] }
\label{inductive}
\end{equation}
where the subscripts denote the relevant frequency modes. It is
straightforward to define the capacitance matrix for the circuit
in the same manner:
\begin{equation}
\matrix{
    & \left(\matrix{
        V_{0}^{'}\cr
        V_{0}^{''}\cr
        V_{+}\cr
        V_{-}\cr
    }\right)
    = (-j \omega)^{-1} \left(\matrix{
        C_{A}^{-1} & 0 & 0 & 0 \cr
        0 & C_{B}^{-1} & 0 & 0 \cr
        0 & 0 & C_{C}^{-1} & 0 \cr
        0 & 0 & 0 & -C_{C}^{-1} \cr
    }\right)
        \left(\matrix{
        I_{0}^{'}\cr
        I_{0}^{''}\cr
        I_{+}\cr
        I_{-}\cr
    }\right) \cr
    \cr
\Rightarrow & V[\omega] = Z_{\rm cap} I [\omega] }
\label{capacitive}
\end{equation}
The extra negative sign in $s_{44}$ of Eq. (\ref{capacitive})
accounts for the generation of the conjugate wave amplitude
$a^{\dagger}[\omega_{-}]$ as a result of mixing of carrier
$\omega_{c}$ and signal $\omega_{0}$. Hence, on taking the Fourier
transform of the current and voltage vectors, an extra negative
sign appears for the corresponding coefficient in $Z_{\rm cap}$.
The total $Z$ matrix can then be written as
\begin{equation}
    Z = Z_{\rm ind} + Z_{\rm cap},
\end{equation}
which gives
\begin{equation}
   Z = \left(\matrix{
   -i \delta_{0} Z_{A,B} & 0 & -i \alpha Z_{C} e^{-i\phi} & -i \alpha Z_{C} e^{i\phi} \cr
   0 & -i \delta_{0} Z_{A,B} & \alpha Z_{C} e^{-i\phi} & - \alpha Z_{C} e^{i \phi} \cr
   -i \alpha Z_{A,B} e^{i\phi} & - \alpha Z_{A,B} e^{i\phi} & -i \delta_{\pm}Z_{C} & 0 \cr
   i \alpha Z_{A,B} e^{-i\phi} & -\alpha Z_{A,B} e^{-i\phi} & 0 & -i \delta_{\pm}Z_{C}
   }\right).
   \label{Zmatrix}
\end{equation}
Here we have introduced the symbols
\begin{eqnarray*}
    & & \delta_{0} = \frac{|\omega_{0} -
    \omega_{A,B}|}{\Gamma_{A,B}}; \quad \Gamma_{A,B} = \frac{Z_{A,B}}{2 L_{A,B}}\\
    & & \delta_{\pm} = \frac{|\omega_{\pm} -
    \omega_{C}|}{\Gamma_{C}}; \quad \Gamma_{C} = \frac{Z_{C}}{2 L_{C}}\\
    & & \alpha = \frac{M_{0}}{\sqrt{L_{A,B}L_{C}}},
\end{eqnarray*}
with $\Gamma_{i}$ denoting the linewidth of the $i^{\rm th}$
resonator. On using Eqs. (\ref{Zmatrix}) and (\ref{imp}) in Eq.
(\ref{ZtoS}), we obtain
\begin{eqnarray}
    \left(\matrix{
        a_{0}^{' \rm out}\cr
        a_{0}^{'' \rm out}\cr
        b_{+}^{\rm out}\cr
        b_{-}^{\dagger\rm out }\cr
    }\right)
    &=& \left(\matrix{
    r_{0} & -q_{0} & t_{d} e^{-i \phi} & s_{d} e^{i \phi} \cr
    q_{0} & r_{0} & i t_{d} e^{-i \phi} & -i s_{d} e^{i \phi} \cr
    t_{u} e^{i \phi} & -i t_{u} e^{i\phi} & r_{+} & 0 \cr
    -s_{u} e^{-i \phi} & - i s_{u} e^{-i\phi} & 0 & r_{-}
    }\right)
    \left(\matrix{
        a_{0}^{'\rm in}\cr
        a_{0}^{'' \rm in}\cr
        b_{+}^{\rm in}\cr
        b_{-}^{\dagger \rm in}\cr
    }\right)\\
    &=& S
    \left(\matrix{
        a_{0}^{'\rm in}\cr
        a_{0}^{'' \rm in}\cr
        b_{+}^{\rm in}\cr
        b_{-}^{\dagger\rm in}\cr
    }\right),
    \label{scattmat}
\end{eqnarray}
where $a_{i}^{\rm in/out}$ denotes the shorthand notation for
$a[\omega_{i}]^{\rm in/out}$ [cf. Eq. (\ref{Srelation})]. The
symbols $a, b$ denote the wave amplitudes for left and right ports
respectively. The detailed expressions for scattering coefficients
are listed below:
\begin{eqnarray}
    r_{0} &=& \frac{(\delta_{0}^2+1)(\delta_{\pm} + i)^2 - 4 \alpha^4}{(\delta_{0} + i)^2(\delta_{\pm} + i)^2 - 4 \alpha^4};
    \quad
    q_{0} = \frac{4 \alpha^2 (\delta_{\pm} + i)}{(\delta_{0} + i)^2(\delta_{\pm} + i)^2 - 4 \alpha^4 };
    \label{scattcoeff1}\\
    r_{+} &=& \frac{(\delta_{0} + i)(\delta_{\pm} - i) - 2\alpha^{2}}{(\delta_{0} + i)(\delta_{\pm} +i) - 2 \alpha^2 };
    \quad
    r_{-} = \frac{(\delta_{0} + i)(\delta_{\pm} - i) + 2\alpha^{2}}{(\delta_{0} + i)(\delta_{\pm} + i) + 2 \alpha^2 };
    \\
    t_{u} &=& i \left(\frac{Z_{C}}{Z_{A,B}}\right) \left(\frac{2 \alpha}{(\delta_{0} + i)(\delta_{\pm} + i) - 2 \alpha^2}\right);
    \\
    s_{u} &=& i \left(\frac{Z_{C}}{Z_{A,B}}\right) \left(\frac{2 \alpha}{(\delta_{0} + i)(\delta_{\pm} + i) + 2 \alpha^2}\right);
    \\
    t_{d} &=& i \left(\frac{Z_{A,B}}{Z_{C}}\right) \left(\frac{2 \alpha}{(\delta_{0} + i)(\delta_{\pm} + i) - 2
    \alpha^2}\right);
    \\
    s_{d} &=& i \left(\frac{Z_{A,B}}{Z_{C}}\right) \left(\frac{2 \alpha}{(\delta_{0} + i)(\delta_{\pm} + i) + 2 \alpha^2}\right).
    \label{scattcoeff2}
\end{eqnarray}
Equations (\ref{scattcoeff1})-(\ref{scattcoeff2}) show that the
effective coupling strength $\alpha$ plays the role of the
parametric drive (``pump'') in the UDC. In the limit $\alpha=0$,
transmission coefficients $t_{i}$ and $s_{i}$ become identically
zero while the reflection coefficients $r_{i}$ reduce to those for
three independent series LCR circuits with resonance frequencies
$\omega_{0}$ and $\omega_{c}$ respectively.

We note that the matrix in Eq. (\ref{scattmat}) is one of the
block diagonals of the full $8\times8$ scattering matrix that
describes the interaction of both the positive and negative
frequency wave amplitudes for all participating modes:
\begin{equation}
    \left(\matrix{
    a_{1}^{\rm out} \cr
    a_{1}^{\dagger \rm out} \cr
    a_{2}^{\rm out} \cr
    a_{2}^{\dagger \rm out} \cr
    \vdots \cr
    } \right)
    =  \tilde{S}
    \left(\matrix{
    s_{11} & 0 & \ldots & \ldots \cr
    0 & s_{11}^{*} & \ldots & \ldots \cr
    \vdots & \vdots & \ddots & \vdots \cr
    s_{2N,1} & \ldots & \ldots & s_{2N,2N}
    }\right)
    \left(\matrix{
    a_{1}^{\rm in} \cr
    a_{1}^{\dagger \rm in} \cr
    \vdots \cr
    a_{2N}^{\dagger \rm in}
    } \right).
\end{equation}
This is especially important in the case of an active network
where a conjugation operation is possible i.e. $a_{i} \mapsto
a_{i}^{\dagger} =a [-\omega_{i}]$. The full scattering matrix
$\tilde{S}$ satisfies the following general properties:
\begin{enumerate}
\item det($\tilde{S}$) =1  (as det ($S$) =1) \item $\sum_{j =
1}^{4} |s_{ij}|^2 = 1$ \item $\tilde{S}^{T} J \tilde{S} = J$,
\end{enumerate}
where $J$ represents a symplectic structure defined on the $2N
\times 2N$ phase space ($N$ = number of degrees of freedom),
\begin{equation}
    J = i\sigma_{Y}\otimes I_{N}.
\end{equation}
The last condition of symplecticity follows from the fact that a
transformation of the modes as performed by the scattering matrix
needs to be a canonical transformation. This requirement
translates into the condition for preservation of phase space
volume (information) or the number of participating modes (degrees
of freedom) in the system.
\paragraph*{Derivation of transfer matrix.}
In this section, we derive the transfer matrix for the UDC stage.
This description is equivalent to the usual $ABCD$ matrix of the
circuit theory defined in terms of the voltages and currents for a
two-port network \cite{Pozar},
\begin{equation}
    \left(\matrix{
    V_{b} \cr
    I_{b}
    }\right)
    =
    \left(\matrix{
    A & B \cr
    C & D
    }\right)
    \left(\matrix{
    V_{a} \cr
    I_{a}
    }\right).
\end{equation}
As there exists a straightforward mapping between the reduced wave
amplitudes $a_{i}$ introduced earlier, and the currents and
voltages at the ports \cite{ArchanaDPA},
\begin{eqnarray}
    a_{i}^{\rm out} &=& \frac{V_{i} + Z_{0} I_{i}}{\sqrt{ 2 Z_{0} \hbar \omega_{i}}}\\
    a_{i}^{\rm in} &=& \frac{V_{i} - Z_{0} I_{i}}{\sqrt{ 2 Z_{0} \hbar \omega_{i}}},
\end{eqnarray}
we can easily adapt the concept to the above choice of variables.
\par
For the cascaded chain, we first evaluate the transfer matrix of
each stage and then multiply them to obtain the total transfer
matrix of the cascade. For the UDC stage performing up-conversion,
we can find the relevant transfer matrix by solving Eq.
(\ref{scattmat}) to obtain sideband amplitudes $(a_{+},
a_{-}^{\dagger})$ in terms of low frequency amplitudes
$(a_{0}^{'}, a_{0}^{''})$
\begin{equation}
    \left(\matrix{
        b_{+}^{ \rm out}\cr
        b_{+}^{ \rm in}\cr
        b_{-}^{\dagger \rm out}\cr
        b_{-}^{\dagger \rm in}\cr
    }\right)
    =
    T \left(\matrix{
        a_{0}^{'\rm in}\cr
        a_{0}^{' \rm out}\cr
        a_{0}^{'' \rm in}\cr
        a_{0}^{'' \rm out}\cr
    }\right).
    \label{TUCDC}
\end{equation}
The reversal of `in' and `out' in the column vectors on left and
right hand sides of Eq. (\ref{TUCDC}) is required to maintain a
consistent sense of propagation through the device as the output
of the $(N-1)^{th}$ stage acts as the input for the $N^{th}$ stage
in the chain. On doing the above transformation, we obtain
\begin{equation}
 T_{UC} = \left(\matrix{
         t_{+,LR}e^{i\phi} &  t_{+,RL}^{*} e^{i\phi} & - i t_{+,LR}e^{i\phi} & i t_{+,RL}^{*}e^{i\phi}\cr
         t_{+,RL} e^{i\phi} & t_{+,LR}^{*}e^{i\phi} & -i t_{+,RL} e^{i\phi} & i t_{+,LR}^{*}e^{i\phi}\cr
        t_{-,LR}e^{-i\phi} & t_{-,RL}^{*} e^{-i\phi} & i t_{-,LR}e^{-i\phi} & i t_{-,RL}^{*} e^{-i\phi} \cr
        t_{-,RL} e^{-i\phi} & t_{-,LR}^{*} e^{-i\phi} & i t_{-,RL} e^{-i\phi} &  i t_{-,LR}^{*} e^{-i\phi}
    }\right),
    \label{tuc}
\end{equation}
with
\begin{eqnarray*}
    & & t_{+,LR} = i \left(\frac{Z_{A,B}}{Z_{C}}\right)\left(\frac{(\delta_{0} - i)(\delta_{\pm} - i) -
    2\alpha^2}{4\alpha}\right)\\
    & & t_{+,RL} = i \left(\frac{Z_{A,B}}{Z_{C}}\right)\left(\frac{(\delta_{0} - i)(\delta_{\pm} + i) -
    2\alpha^2}{4\alpha}\right)\\
    & & t_{-,LR} = i \left(\frac{Z_{A,B}}{Z_{C}}\right)\left(\frac{(\delta_{0} - i)(\delta_{\pm} - i)
    + 2\alpha^2}{4\alpha}\right)\\
    & & t_{-,RL} = i \left(\frac{Z_{A,B}}{Z_{C}}\right)\left(\frac{(\delta_{0} - i)(\delta_{\pm} + i)
    + 2\alpha^2}{4\alpha}\right).
\end{eqnarray*}
The subscripts (+,-) refer to the resultant sideband generated at
the output ($\omega_{+}$ or $\omega_{-}^{*}$) while $LR (RL)$
indicates the relevant direction of propagation as
\emph{left-to-right} (\emph{right-to-left}). We note that the
condition for the transformation describing the mapping between
$S$ and $T$ matrices to be non-singular is
\begin{equation}
s_{13}s_{24} - s_{23} s_{14} \neq 0.
\end{equation}
This condition is violated when the couplings $M_{1}$, $M_{2}$
(Fig. S1) are in phase and hence a transfer matrix cannot be
defined in such a case.
\par
Similarly, the scattering matrix of the phase shifting stage,
\begin{equation}
    PS = \left(\matrix{
    0 & 0 & e^{-i\theta} & 0 \cr
    0 & 0 & 0 & e^{i\theta} \cr
    e^{-i\theta} & 0 & 0 & 0 \cr
    0 & e^{i\theta} & 0 & 0
    }\right),
\end{equation}
yields a transfer matrix of the form
\begin{equation}
    T_{PS} = \left(\matrix{
    e^{-i\theta} & 0 & 0 & 0\cr
    0 &  e^{i\theta} & 0 & 0\cr
    0 & 0 & e^{i\theta} & 0 \cr
    0  & 0 & 0 & e^{-i\theta}
    }\right).
\end{equation}
Finally, by exploiting the fact that down-conversion is just the
inverse operation of the phenomenon of up-conversion we obtain the
transfer matrix of the second UDC stage as
\begin{equation}
    T_{DC} = F \times T_{UC}^{-1} \times F,
    \label{tdc}
\end{equation}
where $F$ is the flip matrix of the form $F = i\sigma_{X} \otimes
I_{2}$ required to preserve the consistency in labelling the input
and output amplitudes. The forward and backward propagation
diagrams in Fig. 3b of the main text have been calculated using
the matrices $T_{UC}$ [Eq. (\ref{tuc})] and $T_{DC}$ [Eq.
(\ref{tdc})]. Each arm of the propagation diagram notes the net
forward going amplitude at that frequency: for instance, the
amplitudes for first UDC stage performing upconversion in the
forward propagation diagram (green) can be evaluated using the
first and third rows of the matrix in Eq. (\ref{tuc}) with $\phi
=0$. The net forward going amplitude contributed by $a_{0}^{'}$ to
$a_{+}$ can be calculated as $t_{11} - t _{12} = (t_{+,LR} -
t_{+,RL}^{*}) = -i/2^{3/4}$ while that due to $a_{0}^{''}$ is
given by $ t_{13} - t_{14} = (- i t_{+,LR}- it_{+,RL}^{*}) =
-1/2^{3/4}$, evaluated for the optimal parameter values
$\delta_{0} =0$, $\delta_{\pm} = 1/\sqrt{2}$, $\alpha = 1/2^{3/4}$
reported in the main text. For the backward propagation, a similar
exercise using the second and fourth rows of $T_{UC}$ (and
$T_{DC}$) leads to the propagation diagram shown in the lower
panel of Fig. 3b in red. 
\par
The total transfer matrix
\begin{equation}
    T_{\rm total} = T_{DC_{2}}(\phi=\pi/4) \times T_{PS}(\theta=\pi/2) \times
    T_{UC_{1}}(\phi=0)
\end{equation}
establishes the relationship between the low frequency signals on
the left and the right ports of the cascade as
\begin{eqnarray}
    \left(\matrix{
        b_{0}^{' \rm out}\cr
        b_{0}^{' \rm in}\cr
        b_{0}^{'' \rm out}\cr
        b_{0}^{'' \rm in}\cr
    }\right)
    &=& \left(\matrix{
           i & 0 & 0 & 0 \cr
           0 & 0 & 0 & i \cr
           0 & 0 & i & 0 \cr
           0 & -i & 0 & 0
           }\right)
    \left(\matrix{
        a_{0}^{' \rm in}\cr
        a_{0}^{' \rm out}\cr
        a_{0}^{'' \rm in}\cr
        a_{0}^{'' \rm out}\cr
    }\right)
    \label{transfer}\\
    &=&
    T_{\rm total}
    \left(\matrix{
        a_{0}^{' \rm in}\cr
        a_{0}^{' \rm out}\cr
        a_{0}^{'' \rm in}\cr
        a_{0}^{'' \rm out}\cr
    }\right).
\end{eqnarray}
The matrix in Eq. (\ref{transfer}) has been evaluated for the
resonant case $\delta_{0} = 0$ and the parameter values
\begin{equation}
    \delta_{\pm} = \frac{1}{\sqrt{2}}; \;\;\; \alpha =\frac{1}{2^{3/4}}.
\end{equation}
On evaluating the total scattering matrix $S_{\rm total}$ from
$T_{\rm total}$, we obtain the four port circulator reported in
Eq. (9) of the main text. The device behaves nonreciprocally since
$S^{T} \neq S$ \cite{Pozar}.
\paragraph*{Modulation ellipse.}
This scheme, providing a geometrical visualization of the action
of our device, represents the superposition of two sinusoidal
signals as an ellipse in the plane defined by the quadratures $I$
and $Q$. In general for any two complex phasors $\vec{a}$ and
$\vec{b}$ rotating in opposite directions
\begin{eqnarray}
    \vec{a}e^{i\omega t} + \vec{b}e^{-i\omega t}
    & = & \underbrace{{\rm Re}[(\vec{a} + \vec{b}^{*})e^{i\omega t}]}_{I} \;+\; i \underbrace{{\rm Im}[(\vec{a} - \vec{b}^{*})e^{i\omega t}]}_{Q}.
    \label{IQ}
\end{eqnarray}
The magnitudes and phases of the two complex signals (4 quantities
in total) are encoded as different properties of a colored ellipse
in the $IQ$ plane: the semi-major axis of the ellipse equals
$\rho_{+} = |a| + |b|$ while the semi minor axis equals $\rho_{-}
= \left||a| - |b|\right|$, the angle with the $I$ axis equals
$(\theta_{a} - \theta_{b})/2$ and the location of the colors on
the ellipse represents the phase angle $(\theta_{a} +
\theta_{b})/2$.
\par
The output at each stage in the active circulator comprises two
modes -- the spatially orthogonal modes $(a_{0}^{'}, a_{0}^{''})$
and the sidebands $(a_{+}, a_{-})$ that can be represented as
phasors with opposite sense of rotations in a frame rotating at
the carrier frequency $\omega_{c}$. Thus we can faithfully map the
detailed dynamics of the device after each stage using the
modulation ellipse representation and plot the combined output
signal in the $IQ$ plane with $I = {\rm Re}[(a+b) e^{i \omega t}]$
and $Q= {\rm Im}[(a-b) e^{i \omega t}]$ ($a$ and $b$ representing
the relevant modes at each stage).
\par
Now we present examples of two different kinds of phase rotations
and the resultant transformations (``rotations") of the modulation
ellipse (see Fig. S3 for additional examples on modulation ellipse
representation).
\begin{enumerate}
\item Phase shift:
\\
The action of a phase shifter which performs frequency independent
phase rotations of both the phasors can be described using the
transformations:
\begin{equation}
    a \mapsto a e^{i\theta} \quad {\rm and} \quad b \mapsto b
    e^{i\theta}.
\end{equation}
On using the above and performing the analysis in the $IQ$ plane,
we obtain the expressions for new coordinates as
\begin{eqnarray}
    I = {\rm Re}[a e^{i (\omega t + \theta)} + b^{*}e^{i (\omega t - \theta)}]\\
    Q = {\rm Im}[a e^{i (\omega t + \theta)} - b^{*}e^{i (\omega t - \theta)}].
\end{eqnarray}
The action of such an operation can be easily visualized using the
modulation ellipse as shown in Fig. S2(e).
\begin{figure}
\centering
  \includegraphics[width=0.8\textwidth]{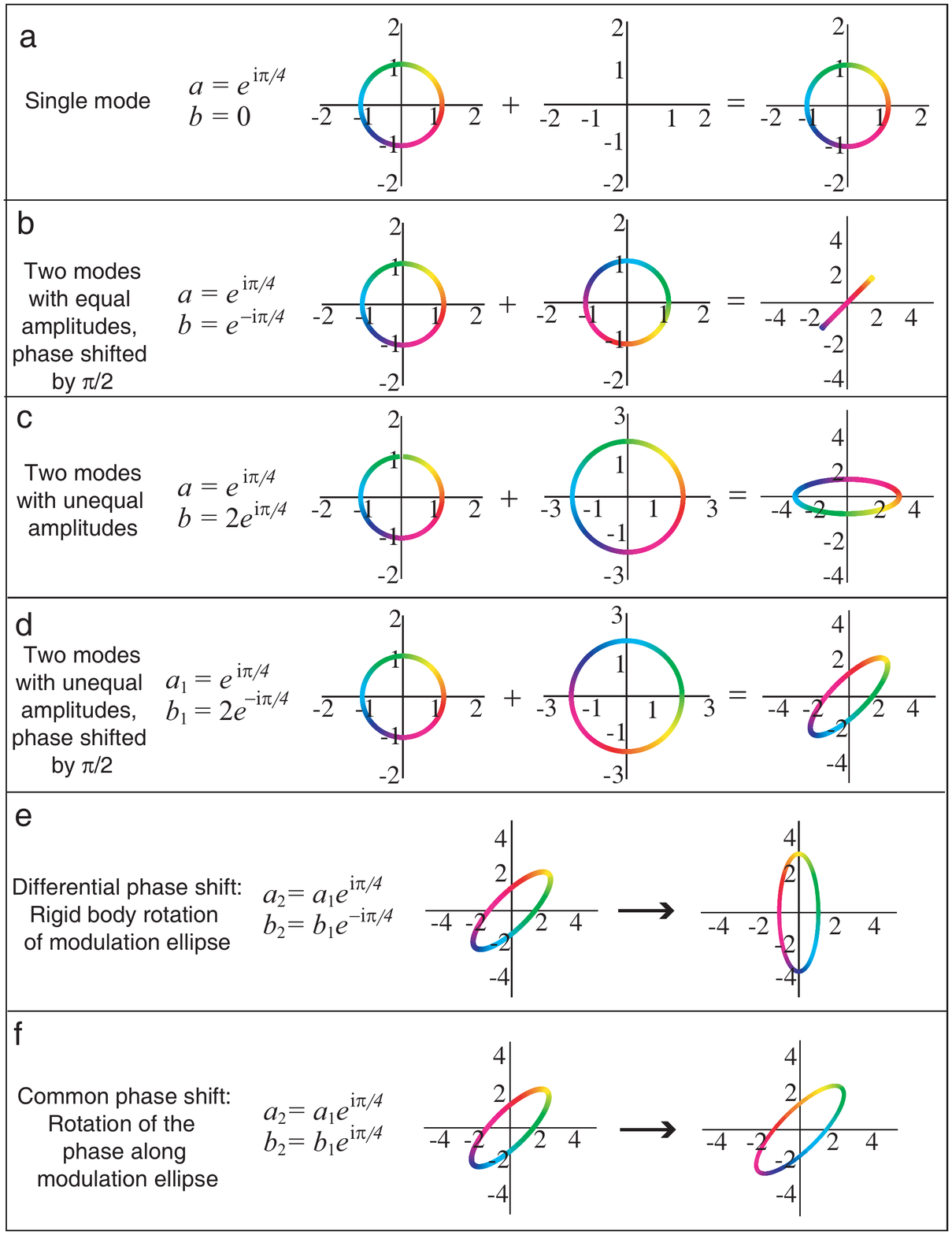}\\
  \caption{{\bf Modulation ellipse representation of two phasors.} In each of the panels, the first column describes the phasors under consideration, the second column
  gives a precise mathematical formula for them and the third column shows the corresponding modulation ellipse. In {\bf a-d}, we show both the input modes and the resultant
  modulation ellipse. The tilt of the ellipse with respect to the $I$ axis represents the relative phase between the two
  modes [$(\theta_{a} - \theta_{b})/2$] while the color along the ellipse represents the average initial phase of the two modes [$(\theta_{a} + \theta_{b})/2$],
  with yellow representing the position of 0 (or $2\pi$). Figs. {\bf(e)} and {\bf(f)} represent the resulting ellipses on performing
 the indicated transformations on the ellipse in {\bf(d)}.}\label{FigS3}
\end{figure}
\item Free evolution:
\\
In contrast to the transformation described above, we now consider
the rotation preformed by a mere time evolution of the two
counter-rotating phasors (say by passage through a transmission
line). In such a case, the phases of the two signals continue to
evolve in opposite directions collecting a phase $\delta$ in time
t ($\delta = \omega t$),
\begin{equation}
    a \mapsto a e^{+i\delta} \quad {\rm and} \quad b \mapsto b e^{-i\delta}.
\end{equation}
The $IQ$ coordinates are calculated as
\begin{eqnarray}
    I = {\rm Re}[(a + b^{*}) e^{i (\omega t + \delta)}] \\
    Q = {\rm Im}[(a - b^{*}) e^{i (\omega t + \delta)}].
\end{eqnarray}
It is immediately evident, that under time evolution, there is
only a trivial phase accumulation leading to change of relative
positions of the two phasors along the circumference of the
modulation ellipse with no rigid rotation of the ellipse, Fig.
S2(f).
\end{enumerate}

In Fig. 3C of the main text, we have used modulation ellipses to
represent the dynamics at each stage of the active circulator. In
the forward propagation direction ($L \;{\rm to}\; R$), the output
ellipse representing ports 3 and 4 has the same orientation as the
input ellipse representing ports 1 and 2; only the average phase
of the total output changes as indicated by the change of color
along the circumference of the output ellipse. In the backward
propagation direction ($R \;{\rm to}\; L$), however, the ellipse
obtained at ports 1 and 2 is rotated by $\pi/2$ with respect to
the input ellipse of ports 3 and 4, in addition to the trivial
average phase change. This indicates the swapping of the
transmitted amplitudes -- port 1 (port 2) receiving the input at
port 4 (port 3) -- that leads to nonreciprocal transmission
characteristics of the device.